\documentclass[english]{article}
\usepackage{spconf,amsmath,graphicx}
\usepackage{amssymb}
\usepackage{amsmath,csquotes}
\usepackage{graphicx}
\usepackage{color}
\usepackage{booktabs}
\usepackage{multirow}
\usepackage{cite}
\usepackage{listings}
\usepackage{subcaption}
\usepackage{booktabs,multirow}
\usepackage[dvipsnames,svgnames,table]{xcolor}
\usepackage{hyperref}
\usepackage{textcomp,gensymb}

\hypersetup{
     colorlinks   = true,
     citecolor    = blue,
     urlcolor     = red 
}

\definecolor{Gray}{gray}{0.90}

\title{Viewport-aware adaptive 360$\degree$ video streaming using tiles \\for virtual reality}
\name{Cagri Ozcinar, Ana De Abreu, and Aljosa Smolic
\address{Trinity College Dublin (TCD), Dublin 2, Ireland.}}

\begin{document}

\maketitle


\begin{abstract}

360$\degree$ video is attracting an increasing amount of attention in the context of Virtual Reality (VR). Owing to its very high-resolution requirements, existing professional streaming services for 360$\degree$ video suffer from severe drawbacks. This paper introduces a novel end-to-end streaming system from encoding to displaying, to transmit 8K resolution 360$\degree$ video and to provide an enhanced VR experience using Head Mounted Displays (HMDs). The main contributions of the proposed system are about tiling, integration of the MPEG-Dynamic Adaptive Streaming over HTTP (DASH) standard, and viewport-aware bitrate level selection. Tiling and adaptive streaming enable the proposed system to deliver very high-resolution 360$\degree$ video at good visual quality. Further, the proposed viewport-aware bitrate assignment selects an optimum DASH representation for each tile in a viewport-aware manner. The quality performance of the proposed system is verified in simulations with varying network bandwidth using realistic view trajectories recorded from user experiments. Our results show that the proposed streaming system compares favorably compared to existing methods in terms of PSNR and SSIM inside the viewport.

\end{abstract}


%
\vspace{-0.8em}

\section{Introduction}
\vspace{-0.8em}

During the last years, significant achievements have been made regarding the rendering capacity and quality of Head Mounted Display (HMD) systems~\cite{Oculusundatedmv}. Modern HMDs can render 360$\degree$ video at a sufficiently high frame-rate and resolution, allowing the viewer to be immersed in the VR environment.

\begin{figure*}[t]
        \centering
        \includegraphics[trim={0.2cm 0.2cm 0cm 0.2cm},clip,width=0.84\textwidth]{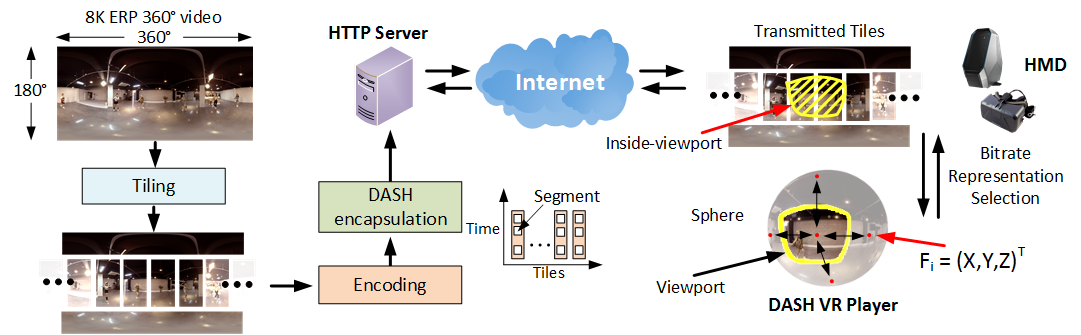}
        \vspace{-0.5em}
        \caption{Schematic diagram of our adaptive 360$\degree$~video streaming system for VR.}
        \vspace{-1.5em}
        \label{mainConcept}
\end{figure*}%

Although numerous professional VR video services have emerged for streaming the 360$\degree$ video, \textit{e.g.,} YouTube 360~\cite{youtube360}, they suffer from severe drawbacks. More clearly, each video frame, containing a 360$\degree$ Field of View (FOV), is encoded and transmitted regardless of the FOV of the HMD. In fact, the existing HMDs have a viewable FOV that ranges from 96$\degree$ to 110$\degree$~\cite{Jtc1sc29wg2016_np}, meaning they use only around one fifth of the transmitted data~\cite{HuaweiTechnologiescoLTD2016mb}. The area of the 360$\degree$~video frame displayed by the HMD at a given time is known as the \emph{viewport}. Hence, the perceptual quality of such video mainly depends on the viewport quality. In existing professional services, regions \emph{outside} the viewport waste a considerable proportion of the bandwidth. This unnecessarily utilized bandwidth results in overall low-quality video streaming to comply with the present Internet and decoding limitations.

Considering its data-intensive representation and the best-effort nature of the Internet, 360$\degree$~video streaming requires a bitrate adaptive solution to offer an enhanced VR experience. To this end, MPEG-Dynamic Adaptive Streaming over HTTP (DASH)~\cite{Isoiec230091undatedlt}, which is an international standard for adaptive streaming, is a key enabler in achieving a smooth VR video playback. The main aim of DASH is to provide a high-quality streaming experience based on the client bandwidth. In DASH, video streams are requested using a manifest file, Media Presentation Description (MPD), which contains a set of bitrate representations.

The significant requirements regarding resolution to ensure high-quality VR experience~\cite{mossVRsickness,Jtc1sc29wg2016_np} can be managed with tiling and viewport-aware solutions using DASH. For example, tiling can generate self-decodable regions, \textit{i.e.,} tiles, by \emph{spatially} dividing the frames.~In addition, tiling can consider the importance of regions~\cite{Meddeb2015_eh} and can enable parallel downloading~\cite{stenberg2014http2} and decoding~\cite{6547985, Niamut2016hd} features.~As the HMDs use only the viewport out of the captured 360$\degree$ video, the visual quality of the viewport can be enhanced in a viewport-aware manner.

This paper focuses on adaptive 360$\degree$ video streaming and provides improved viewport quality compared to existing professional services. A VR video streaming system, from encoding to displaying, is designed to transmit 8K 360$\degree$ video and to offer an enhanced VR experience. The main contribution of this work is an end-to-end streaming system implementation that contains tiling, a novel extension of the MPD, and DASH bitrate level selection in a viewport-aware manner. Experimental results showed that the proposed system demonstrates significant quality enhancements compared with the streaming approach that is currently used by professional VR services~\cite{youtube360,std360industry} in this area. Our work concentrates on adaptive distribution of bitrate and quality over the 360$\degree$ video in contrast to existing professional services. The system uses the tiling concept to divide each video frame into tiles to encode, transmit, and decode them effectively. To facilitate tiled 360$\degree$ video streaming, we extended the concept of DASH Spatial Relationship Description (SRD)~\cite{ISOIEC2300912014Amd22015ye}, and introduced a new MPD for DASH streaming. The proposed DASH player also efficiently distributes the available bandwidth to the tiles and requests the best bitrate representation for each tile in a viewport-aware manner.

The rest of this paper is organized as follows: the related work is presented in Sec.~\ref{releatedWork}. In Sec.~\ref{motivation}, we introduce the proposed solution. In Sec.~\ref{experiment}, we present the experimental results. Finally, Sec.~\ref{conclusion} concludes the paper. 
\vspace{-1.2em}
\section{Related work}
\label{releatedWork}
\vspace{-0.8em}
Several early studies~\cite{Grunheit2002du, Smolic2005rep, 6012244, heymann2005representation} used tiling for the aim of viewport-aware 360$\degree$ video transmission, where an HMD technology was utilized without using a bitrate adaptive streaming.~Although these solutions are far from achieving the expected high-quality performance because of the limitations of the early technologies, they are pioneering works in this area.

Recent advances in adaptive streaming and modern HMDs have made it feasible to deliver and render high-quality 360$\degree$~video. There are two types of adaptive streaming solutions which are commonly used in this area, namely, non-tiled and tiled. In the context of non-tiled, a recent work in~\cite{Corbillon2016js} focused on the quality impact of 360$\degree$~video projections. In addition, in the context of tiling research, a short paper of Skupin \textit{et. al} in~\cite{Frounhoundatedwi} described the bandwidth problem of 360$\degree$~video, and suggested to use tile-based streaming. Also, in the work in~\cite{LeFeuvre2016pg}, high-resolution video content is transmitted in tiled fashion using fixed rectangular tiles, such as 5$\times$5.

To design a practically applicable streaming system for 360$\degree$~video and to address studies needed in this area~\cite{Jtc1sc29wg2016wm}, we developed a system for VR that supports tiled and viewport-aware adaptive streaming. We used unequal tile sizes and viewport-aware bitrate \textit{distribution} using a novel \textit{distance} criterion. To verify our method, we recorded \textit{real} viewport trajectories from subjects in viewing sessions. With using our recorded data, we calculated the viewport quality scores and compared our proposed method with the reference solution, which is based on the existing professional adaptive streaming systems~\cite{std360industry}.

\vspace{-1.2em}
\section{Proposed Streaming System}
\label{motivation}
\vspace{-0.8em}
Typically 360$\degree$ video is shot using a collection of cameras that cover a 360$\degree$ FOV. The captured frames are then stitched together and projected onto a 2D plane using the Equi-Rectangular Projection (ERP)~\cite{projections}. The overall video has to have a very high resolution such as 8K in order to provide high quality for a given viewport, which is only a portion of the overall view.

To increase the streaming performance of such content, we propose an adaptive streaming system that enhances the visual quality of the video displayed in the viewport. The proposed system divides each ERP video frame into self-decodable tiles, encodes them at various bitrates, and then encapsulates and stores them in the HTTP server. Each bitstream contains multiple self-decodable time segments for the purpose of adaptive streaming. 

The proposed DASH VR player requests the most appropriate bitrate representations for each tile given the available network bandwidth and the viewport location. To avoid undesirable quality degradation during a possible sudden viewport movement, our current system streams the whole tiled ERP frame by gradually reducing the bitrate (or quality) of the outside-viewport tiles and increases the quality of the inside-viewport tiles. To this end, in order to use our approach with the DASH standard, we extended the DASH-SRD representation, and contributed a novel MPD for DASH streaming. Finally, the decoded tiles are rendered in the HMD after projecting them back into a sphere. A schematic diagram summarizing the proposed streaming system is presented in Fig.~\ref{mainConcept}.

\vspace{-1em}
\subsection{Tiling and encoding}
\vspace{-0.5em}

Tiling divides video frames into several self-decodable streams (tiles) to deliver and decode high-resolution 360$\degree$ video effectively. Consequently, selection of the tile size plays an important role regarding streaming performance. For instance, using large tiles may increase the compression efficiency, but it may contain many pixels that are not a part of the current viewport. In contrast, using small tiles may reduce the compression efficiency because of exploiting less spatial redundancies. Consequently, in the context of this work, we pay attention to the typically lower importance and low-motion characteristics of the poles and the dominant viewing adjacency of the equator~\cite{Yu2015ox}. First, each ERP video frame is vertically divided into three parts: the equator and two poles. The equator represents the middle segment, and the two poles stand for the top and the bottom sections of the frame. As the poles occupy the largest regions of redundant pixels~\cite{Budagavi2015bv}, in those areas, larger tile resolution size can be used to compress them using a lower bitrate. Additionally, as the equator is associated with the most dominant viewing adjacency~\cite{Yu2015ox}, it is further divided horizontally into several tiles to efficiently transmit them using the proposed viewport-aware representation selection, as shown in Fig.~\ref{mainConcept}.

Each tiled video frame may then be encoded e.g. with either the H.264/Advanced Video Coding~(AVC)~\cite{Isoiec1449610undatedkl} or the High Efficient Video Coding~(HEVC) standard (H.265/MPEG-HEVC). As the proposed system is a codec agnostic, the player can decode each tile stream with either of the coding standards, along with several possible individual decoders for AVC or a single decoder for HEVC encoded tiles. Importantly, tiling offers additionally parallel downloading and decoding opportunities to transmit and decode the high-resolution 360$\degree$ video effectively.

\vspace{-1em}
\subsection{MPD extension for 360$\degree$ video}
\vspace{-0.5em}

In order to transmit the tiled video frames to the VR client, the DASH standard is used for adaptive streaming. To this end, Each encoded tile is divided into self-playable time segments, and an MPD is delivered before starting a streaming session. The MPD describes the structure of bitrate representations for each tile. 


In order to use our streaming approach with the DASH standard, a novel MPD structure is introduced for 360$\degree$ video content. For that, the standard MPD structure for SRD is extended to support our viewport-aware approach. The proposed MPD includes the tiles' Identification Numbers (ID), their resolutions, their positions and their center locations in terms of the spherical Cartesian coordinates. This way, each tile can be requested with its ID and then mapped to a sphere with the help of its resolution and its position on the sphere. To facilitate viewport-aware streaming, a center location is added in the MPD for each tile. This way, each $i^{th}$ tile contains a center point, which is represented as $\textbf{F}_i = (X, Y, Z)^T$, as illustrated in Fig. \ref{mainConcept}.

\vspace{-1em}
\subsection{Viewport-aware representation selection}
\label{viewportaware}
\vspace{-0.5em}

Given the current viewport, an optimal bitrate representation is selected for each tile. This work is interested in providing more importance to the tiles in the viewport, meaning higher bitrate. Consequently, the total bits to be assigned to the tiles that are \emph{outside} of the viewport are reduced. The perceived visual quality is enhanced by increasing the bitrate of the viewport. The bitrate of the tiles outside of the viewport is gradually reduced based on the \emph{distance} between the spherical center location of the viewport and each outside-viewport tile.

The proposed DASH VR player requests the optimum bitrate representation for each tile using the designed MPD. To this end, we define \mbox{\boldmath$V$}, \mbox{\boldmath{$S^{in}$}} and \mbox{\boldmath{$S^{out}$}} for the viewport, a set of the tiles inside the viewport, and a set of tiles outside of the viewport, respectively. The bitrate assigned to the $i^{th}$ tile in the viewport is the following:
\vspace{-0.3em}
\begin{equation} 
\vspace{-0.5em}
\label{eq1}
R_{V_i} = (\gamma R_{cur})\omega_i \quad i\in \mbox{\boldmath$V$}, \quad i \not\in \mbox{\boldmath{$S^{out}$}},
\end{equation}
where the tile represents as $i$, $ i\in\mathbb{Z}$ and $i \in [1, N]$. $N$ denotes the total number of tiles. $\gamma$ is a client-defined constant term, which is $\gamma \in [0, 1]$. In this work, $\gamma$ is selected as 0.8 empirically. $R_{cur}$ is the current available bandwidth, and $\omega_i$ is the weight of the $i^{th}$ tile in \mbox{\boldmath{$S^{in}$}}. This weight is calculated for the $i^{th}$ tile as:
\vspace{-0.5em}
\begin{equation} 
\label{eq2}
\omega_i=\frac{\#\text{ of pixels in }(\mbox{\boldmath$V$} \cap \mbox{$S^{in}_i$})}{\rho_{tot}},
\vspace{-0.2em}
\end{equation}
where $\rho_{tot}$ is the total number of pixels in \mbox{\boldmath$V$}.

To gradually distribute the remaining bandwidth among the outside-viewport tiles, the Euclidean distance, $\delta_i$, is calculated between the middle point of the \mbox{\boldmath$V$}, and each defined $\textbf{F}_i$ point for each outside-viewport tile, \mbox{\boldmath{$S^{out}_i$}}. The bitrate estimation for the $i^{th}$ tile in \mbox{\boldmath{$S^{out}$}} is calculated as follows:
\begin{equation} 
\label{eq11}
R_{\mbox{$S^{out}_i$}}=\widehat{\kappa_i}((1-\gamma) R_{cur}) \quad i \in \mbox{\boldmath{$S^{out}$}}, \quad i \not\in \mbox{\boldmath{$V$}},
\vspace{-0.5em}
\end{equation}
where $\widehat{\kappa_i}$ is calculated as $\widehat{\kappa_i}=\frac{\kappa_i}{\Sigma_i\kappa_i}$ and $\kappa_i=\frac{\max_i \delta_i}{\delta_i}$. 

Finally, the client requests a bitrate representation for each tile, which can be obtained as follows:
\begin{equation} 
J\Leftarrow \min_{J}|r_J-R_{i}|\quad i\in (\mbox{\boldmath{$S^{in}$}} \cup \mbox{\boldmath{$S^{out}$}}),
\end{equation}
where R= $R_{\mbox{$S^{out}$}}$ $\cup$ $R_{\mbox{$S^{in}$}}$, $J$ is the selected DASH representation ID, $J\in \{1,\dots,\epsilon\}$ given $\epsilon$ is the total number of representations, and $r_J$ is the bitrate of the $J^{th}$ representation. 
\vspace{-1.3em}
\section{Experiments}
\label{experiment}

\begin{figure*}[htbp]
        \centering
        \begin{subfigure}[b]{0.245\linewidth}
                \centering
                \includegraphics[trim=0cm 0cm 0cm 0cm, clip=true, width=\linewidth,draft=false]{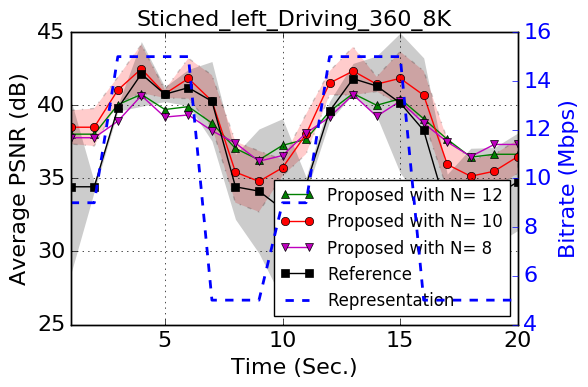}
                \vspace{-1.3em}
        \end{subfigure}
        \begin{subfigure}[b]{0.245\linewidth}        
                \centering
                \includegraphics[trim=0cm 0cm 0cm 0cm, clip=true, width=\linewidth,draft=false]{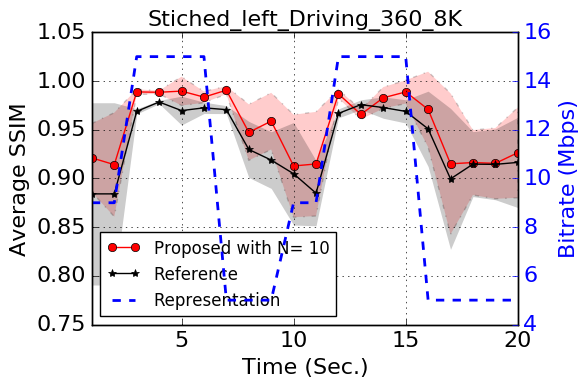}
                \vspace{-1.3em}
        \end{subfigure}
        \begin{subfigure}[b]{0.245\linewidth}
                \centering
                \includegraphics[trim=0cm 0cm 0cm 0cm, clip=true, width=\linewidth,draft=false]{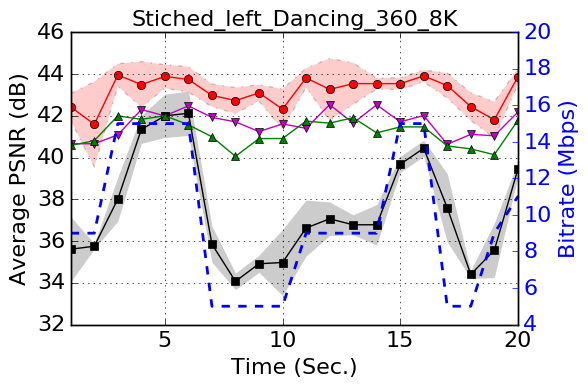}
                \vspace{-1.3em}
                \label{fig:res:p:1:2}
        \end{subfigure}
        \begin{subfigure}[b]{0.245\linewidth}        
                \centering
                \includegraphics[trim=0cm 0cm 0cm 0cm, clip=true, width=\linewidth,draft=false]{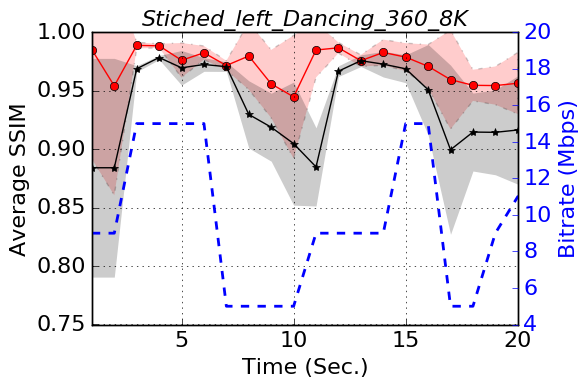}
                \vspace{-1.3em}
                \label{fig:res:p:1:4}               
        \end{subfigure}
        \vspace{-1em}
        \caption{Quality over time comparison between the proposed and the reference adaptive streaming solutions.}
        \vspace{-1.5em}
        \label{per}   
\end{figure*}

\vspace{-1em}
\subsection{Setup}
\vspace{-0.3em}

Two 360$\degree$~video sequences from the MPEG video exploration experiments~\cite{2016rr} were used: \textit{Stiched\_left\_Driving360\_8K} and \textit{Stiched\_left\_Dancing360\_8K}. Their resolutions are 8K, 8192$\times$4096. We focused on the browser-based VR use-case that is one of the core experiments in the ongoing standardization activity for this subject~\cite{Jtc1sc29wg2016wm}. Since AVC is the only implemented decoder in current browsers that handle HMDs, AVC encoded streams were tested over a real network. To this end, the x264 software (\textit{ver.} r2643)~\cite{x264} was used for encoding purposes. We used 0.9, 2, 5, 7, 9, 11, 13, 15, 17, 19, 21, 23, and 25 \textit{Mbps} as \textit{target bitrates} for the proposed and reference methods. The MP4Box~\cite{mp4box} was used to wrap the encoded content within an MP4 header file. Then this tool was utilized to create 2 \textit{sec.} time segments for each tile. Also, a server-client test-bed was implemented to analyze the performance of the proposed system in a realistic network environment. We evaluated the algorithms at a perceptually acceptable quality level for VR video applications~\cite{HuaweiTechnologiescoLTD2016mb}. The connection between the server and the client was a local wired connection, with varying bandwidth between 4 and 22~\textit{Mbps}. 

The proposed method, denoted as \textit{proposed}, divides each video frame into $N$ tiles. Two tiles were utilized for the poles, and $N-2$ tiles were used for the equator. In this test, we experimented with several settings of $N$. Encoded bitrate for each tile is equally distributed by dividing the \textit{target bitrate} to the $N$ tiles. The DASH VR player was implemented using three APIs, namely, three.js~\cite{threeJS}, WebVR~\cite{webvr}, and dash.js~\cite{dashJS}. On the contrary, the reference system, denoted as \textit{reference}, which works in similar principle in existing professional streaming services, neither uses viewport-aware nor tiling techniques. The \textit{reference} transmits each 8K ERP video frame as a single tile using DASH. 


PSNR and SSIM measures were employed to evaluate the objective viewport quality. For that, we recorded real view trajectories of 8 users viewing the content on Oculus Rift DK2~\cite{Oculusundatedmv}. Each participant session started after a 10 \textit{sec.} training video. Average inside-viewport quality scores were calculated over time using the participants' trajectories. 


\vspace{-0.9em}
\subsection{Performance Evaluation}
\vspace{-0.3em}

To evaluate the quality performance of the \textit{proposed} and the \textit{reference} adaptive streaming methods over varying network throughput, average qualities were calculated using the participants' changing viewport over the time. $N$ = 12, 10 and 8 were tested to investigate the impact of the number of tiles. Fig. \ref{per} shows the comparison between the \textit{proposed} and the \textit{reference} streaming solutions in terms of \textit{average} PSNR and SSIM with their \textit{variations} over time. In the figure, the filled area shows the variation, and the left and right axes represent the objective measures and the bitrate representation, respectively. The dashed-line denotes the selected bitrate representation by the DASH client. 

\begin{figure}[h]
\centering
\includegraphics[width=0.80\linewidth]{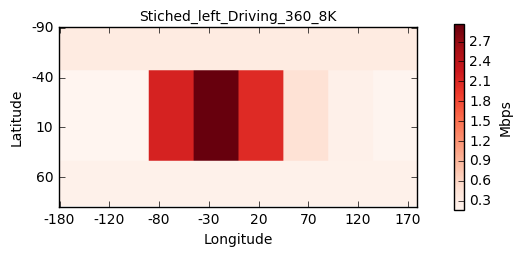}
\vspace{-1.2em}    
\caption{Bitrate distribution of the $4^{th}$ participant at 4 \textit{sec.} with $N$ = 10. Selected representation and calculated viewport quality are 9 \textit{Mbps} and  40.156 \textit{dB}, respectively.}
\label{bitrateDist}
\vspace{-1.2em}    
\end{figure}

The results show that the \textit{proposed} with $N$=10 considerably increases the streaming quality compared with the \textit{reference} at all times.~Also, the results show a quality drop after each 1~\textit{sec.}~segment playback.~The reason is the viewport movement before delivering new segments. Because we used a distance-based bitrate distribution, visual quality was preserved effectively, compared with the \textit{reference}, for each content. The bitrate distribution of the \textit{Stiched\_left\_Driving360\_8K} sequence is illustrated in Fig \ref{bitrateDist}, where the inside-viewport tiles use the highest bitrates. In addition, as it can be seen in Fig \ref{per}, the \textit{proposed} with $N$=10 achieves 1.66/0.02 and 5.72/0.04 average PSNR(dB)/SSIM gains with respect to the \textit{reference} method for the \textit{Stiched\_left\_Driving360\_8K} and \textit{Stiched\_left\_Dancing\\360\_8K} sequences, respectively. 
                



In this work, we demonstrated that the \textit{proposed} streaming performance depends on the viewport movement activity and content type. For example, for the \textit{Stiched\_left\_Driving360\\\_8K} sequence, the \textit{proposed} with $N$ = 12 and 8 methods preserves complex details at low bitrate, and provides higher visual quality. Also, in the \textit{proposed} with $N$ = $\left\{12, 10, 8\right\}$ methods, a considerable quality gain was achieved with respective to the \textit{proposed} for the \textit{Stiched\_left\_Dancing360\_8K} sequence. In this sequence, the \textit{proposed} with $N$ = 10, which uses the optimum $N$ size in our system, achieved the highest quality performance relative to using $N$ = $\left\{12, 8\right\}$.

\vspace{-1.2em}
\section{Conclusions}
\label{conclusion}
\vspace{-0.8em}
This paper introduced a novel end-to-end streaming system for VR, which resulted in enhanced viewport quality under varying bandwidth and different viewport trajectories. The proposed system includes tiling, a novel MPD for DASH, and viewport-aware bitrate level selection methods. The quality performance of the proposed system was verified in simulations with varying network bandwidth using realistic view trajectories recorded from user experiments. Experimental results showed that significant quality enhancement was achieved by the proposed method compared with the reference solution that is currently used by professional VR streaming services. Future research will be devoted to investigating tile size optimization and tile discarding, as they are expected to further increase the visual quality. 





\bibliographystyle{IEEEbib}
\bibliography{main}




\end{document}